# Polarization Multiplexed Diffractive Computing: All-Optical Implementation of a Group of Linear Transformations Through a Polarization-Encoded Diffractive Network

Jingxi Li[1,2,3], Yi-Chun Hung[1], Onur Kulce[1,2,3], Deniz Mengu[1,2,3], and Aydogan Ozcan[1,2,3*]

[1]Electrical and Computer Engineering Department, University of California, Los Angeles, CA, 90095, USA

[2]Bioengineering Department, University of California, Los Angeles, CA, 90095, USA

[3]California NanoSystems Institute (CNSI), University of California, Los Angeles, CA, 90095, USA

[*]Correspondence to: ozcan@ucla.edu

**Abstract:**

Research on optical computing has recently attracted significant attention due to the transformative advances in machine learning. Among different approaches, diffractive optical networks composed of spatially-engineered transmissive surfaces have been demonstrated for all-optical statistical inference and performing arbitrary linear transformations using passive, free-space optical layers. Here, we introduce a polarization multiplexed diffractive processor to all-optically perform multiple, arbitrarily-selected linear transformations through a single diffractive network trained using deep learning. In this framework, an array of pre-selected linear polarizers is positioned between trainable transmissive diffractive materials that are isotropic, and different target linear transformations (complex-valued) are uniquely assigned to different combinations of input/output polarization states. The transmission layers of this polarization multiplexed diffractive network are trained and optimized via deep learning and error-backpropagation by using thousands of examples of the input/output fields corresponding to each one of the complex-valued linear transformations assigned to different input/output polarization combinations. Our results and analysis reveal that a single diffractive network can successfully approximate and all-optically implement a group of arbitrarily-selected target transformations with a negligible error when the number of trainable diffractive features/neurons ($N$) approaches $N_p N_i N_o$, where $N_i$ and $N_o$ represent the number of pixels at the input and output fields-of-view, respectively, and $N_p$ refers to the number of unique linear transformations assigned to different input/output polarization combinations. This polarization-multiplexed all-optical diffractive processor can find various applications in optical computing and polarization-based machine vision tasks.

**Introduction**

With the increasing global demand for machine learning and computing in general, using light to perform computation has been a rapidly growing focus area of optics and photonics[1–5]. The research on optical computing has a long history spanning decades of exciting research and development efforts[6–31]. Motivated by the massive success of artificial intelligence and deep learning, in specific, a myriad of new hardware designs for optical computing have been reported recently, including, e.g., on-chip integrated photonic circuits[16–22], free-space optical platforms[23–28], and others[29–31]. Among these different optical computing systems, the integration of successive transmissive diffractive layers (forming an optical network) has been demonstrated for optical information processing, such as object classification[23,32–43], image reconstruction[38,44], all-optical phase recovery and quantitative phase imaging[45], and logic operations[46–48]. A diffractive network is trained using deep learning and error-backpropagation methods implemented in a digital computer, after which the resulting transmissive layers are fabricated to form a physical network that computes based on the diffraction of the input light through these spatially-engineered transmissive layers. Because the computational task is completed as the light passes through thin and passive optical elements, this approach is very fast, and the inference process does not consume power except for the illumination light. It is also scalable since an increase in the input field-of-view (FOV) can be handled by fabricating larger transmissive layers and/or deeper diffractive designs with more successive layers positioned one after another. Furthermore, both the phase and the amplitude information channels of the input scene/FOV can be processed by a diffractive optical network, without the need for phase retrieval or digitizing, vectorizing an image of the scene, which makes diffractive computing highly desirable for machine vision applications[38,44]. Harnessing light-matter interactions using engineered diffractive surfaces also enabled the inverse design of optical elements for e.g., spatially-controlled wavelength demultiplexing[49], pulse engineering[50], and orbital angular momentum multiplexing/demultiplexing[51,52]. It has also been shown that a diffractive network can be trained by optimizing its diffractive layers to perform an arbitrary complex-valued linear transformation between its input and output fields-of-view, demonstrating its computing capability for complex-valued matrix-vector operations at the speed of light propagation through a passive diffractive system.

All these results highlight the unique capabilities of diffractive networks to manipulate various physical properties of light, including e.g., its amplitude and phase distribution, spatial frequency, spectral bandwidth, orbital angular momentum, for performing specific computational tasks that are desired. As another important physical property of light, polarization specifies the geometrical orientation of electromagnetic wave oscillations. Utilizing the polarization state of light has played a pivotal role in numerous applications, including telecommunications[53–55], imaging[56–61], sensing[62–64], computing[65], and displays[66,67]. For example, polarization-division multiplexing (PDM) has been used in telecommunication systems to permit two channels of information to be simultaneously transmitted using orthogonal polarization states over a single wavelength[54,68].

Here, we report the design of polarization-multiplexed diffractive optical networks to perform a group of arbitrary linear transformations using a common set of diffractive layers that are jointly-optimized to all-optically perform each one of the target complex-valued linear transformations at a different combination of input/output polarization states. In our earlier work[69], we showed that a diffractive optical network composed of spatially-engineered layers could all-optically perform an arbitrary complex-valued linear transformation between an input and output field-of-view with a negligible error when the number of trainable diffractive elements/neurons ($N$) approaches $N_i N_o$, where $N_i$ and $N_o$ represent the number of pixels at the input and output FOVs, respectively. In this work, we use polarization multiplexing between the input and output FOVs of a diffractive network to increase the capacity of diffractive computing and all-optically perform a group of arbitrary linear transformations that are complex-valued. These polarization multiplexed diffractive network designs are *not* based on birefringent, anisotropic or polarization-sensitive materials; instead, our designs utilize standard diffractive surfaces where the phase and amplitude transmission coefficients of each trainable diffractive feature are independent of the polarization state of the input light. Using a network design solely based on standard isotropic diffractive materials makes our designs simpler in terms of material selection, fabrication and scale-up; however, it also makes the diffractive network insensitive to different polarization states, and therefore, polarization multiplexed all-optical computation of different transformations becomes impossible. To overcome this challenge, we used a non-trainable, pre-determined array of linear polarizers (at 0°, 45°, 90° and 135°) within the diffractive network that acted as polarization seeds for the trainable isotropic diffractive layers to all-optically execute different linear transformations through input/output polarization multiplexing (see Fig. 1a). Stated differently, we used data-driven training and optimization of isotropic diffractive layers to encode different linear transformations into different input/output polarization combinations, and this encoding is made possible by the polarization mode diversity introduced by a non-trainable, pre-determined array of linear polarizers within the diffractive volume.

In our first implementation, we performed two different, arbitrarily-selected linear transformations (i.e., $N_p = 2$) using a diffractive network composed of four transmissive layers that are jointly optimized using deep learning, where the first target linear transformation was assigned to x (0°) linear input and x linear output polarization combination, and the second target linear transformation was assigned to y (90°) linear input and y linear output polarization combination. For this case of $N_p = 2$, there are two different schemes (Fig. 1b) to all-optically access/implement the desired linear transformations: sequential (x and y input polarization states encode the input information sequentially, one after another) or simultaneous (x and y input polarizations encode the input information at the same time within the input FOV). Our numerical results (Figs. 2-5) reveal that one can successfully train a diffractive network under each one of these operation modes (sequential vs. simultaneous) to approximate the two target, arbitrary-selected linear transformations with a negligible error when the number of trainable diffractive neurons $N$ approaches $N_p N_i N_o = 2 N_i N_o$.

In our second implementation (Fig. 6), we performed four different, arbitrary linear transformations (i.e., $N_p = 4$) using a diffractive network composed of eight transmissive layers that are jointly optimized using deep learning and examples of input/output fields corresponding to the selected complex-valued linear transformations (ground truth). In this case, the first target transformation was assigned to x linear input and 45° linear output polarization combination, the second target transformation was assigned to y linear input and 135° linear output polarization combination, the third target transformation was assigned to x linear input and 135° linear output polarization combination and finally the fourth target transformation was assigned to y linear input and 45° linear output polarization combination. Our analyses of this 4-channel polarization multiplexed diffractive system show that when $N \geq N_p N_i N_o = 4 N_i N_o$, all the target linear transformations can be successfully approximated, following a similar conclusion as in the first implementation case ($N_p = 2$).

Without the use of a non-trainable, pre-determined array of linear polarizers acting as polarization seeds within the network, none of these multiplexing results could be achieved using isotropic diffractive materials, no matter how they are trained or optimized, since they would normally perform the same transformation under different input polarization states.

Our results should *not* be confused with polarization multiplexed (or wavelength/illumination multiplexed) projection of a set of desired complex fields at the output of a metamaterial design; such multiplexed metamaterial systems do not implement an arbitrary matrix multiplication operation. Each input-output polarization combination in our diffractive design represents an all-optical implementation of a unique linear transformation between the input and output FOVs. Therefore, for each input-output polarization combination, infinitely many different target complex fields can be all-optically synthesized by the trained diffractive network in response to different input field distributions; and this capability accurately defines the corresponding complex-valued linear transformation at the output FOV for all the possible and infinitely many combinations of phase and amplitude distributions at the input FOV.

A polarization multiplexed diffractive network can perform an arbitrary set of target linear transformations using the same diffractive layers that all-optically implement a distinct complex-valued linear transformation at a selected input/output polarization combination. We believe that this unique framework will be valuable in developing high-throughput optical processors and polarization-based machine vision systems operating at different parts of the electromagnetic spectrum. Moreover, the presented diffractive computing platform and the underlying concepts can be used to develop polarization-aware optical information processing systems for e.g., detection, localization, and statistical inference of objects with unique polarization properties.

**Results**

Throughout this section, the terms "diffractive optical network," "diffractive network," and "diffractive processor" are interchangeably used. The schematic of our framework for 2-channel polarization multiplexed all-optical computing ($N_p = 2$) is shown in Fig. 1a. A polarization-encoded diffractive neural network, composed of 4 trainable diffractive layers, is trained to all-optically perform 2 distinct, complex-valued linear transformations between the input and output FOVs through 2 orthogonal polarization channels. The pre-determined polarizer array (which is treated as non-trainable) consists of multiple linear polarizer units with four different polarization directions: 0°, 45°, 90° and 135°, and it is located between the 2$^{nd}$ and 3$^{rd}$ trainable diffractive layers. More details about the architecture, optical forward model and training details of the polarization diffractive network can be found in the Methods section.

We use $i$ and $o$ to denote the complex-valued, vectorized versions of the 2D input and output complex fields located at the input and output FOVs of the diffractive network, respectively, as presented in Fig. 1a. Based on the scalar diffraction theory, here $i_x$ and $o_x$ represent the column vectors of the complex fields generated by sampling the x-polarized optical fields within the input and output FOVs, respectively, and vectorizing the resulting 2D matrices in a column-major order. Similar to $i_x$ and $o_x$, $i_y$ and $o_y$ are their counterparts generated by sampling the y-polarized optical fields within the input and output FOVs, respectively. Based on this notation, ($i_x$, $i_y$) and ($o_x$, $o_y$) can be considered to represent the input and output channels of our polarization multiplexed diffractive network, respectively. In our analyses, the number of pixels in the input and output FOVs are both taken as $N_i = N_o = 8^2 = 64$, such that each target linear transformation matrix has $64^2$ complex-valued entries.

In this first implementation with $N_p = 2$, we randomly generated two complex-valued matrices $A_1$ and $A_2$, each with a size of $N_i \times N_o = 64^2$, to serve as two unique arbitrary linear transformations that we would like to all-optically implement using a single polarization diffractive network. Visualized in Fig. 2a with their amplitude and phase components, these two matrices are independently generated using different random seeds, and the difference between the two matrices can be found in Fig. S1. We also randomly generated two training sets of complex-valued vectors $\{i_1\}$ and $\{i_2\}$ with $N_i = 64$ as input fields, and constructed the corresponding sets of output field vectors $\{o_1\}$ and $\{o_2\}$ using $o_1 = A_1 i_1$ and $o_2 = A_2 i_2$, respectively. For each one of these training sets, $\{i_1\}$ and $\{i_2\}$, we used 55,000 randomly generated complex fields in our training process. A further increase in the size of this training dataset (to e.g., >100,000 randomly generated complex fields) could improve the transformation approximation accuracy of the trained diffractive networks, but would not change the general conclusions of this manuscript and therefore is left as future work.

Based on the given inputs of $\{i_1\}$ and $\{i_2\}$, the ultimate goal of training our polarization multiplexed diffractive network is to simultaneously compute the all-optical output fields $\{o'_1\}$ and $\{o'_2\}$ to come close to the output ground truth (target) fields $\{o_1\}$ and $\{o_2\}$; this way, the all-optical transformations $A'_1$ and $A'_2$ performed by the trained single diffractive system represent an accurate approximation to their ground truth (target) transformation matrices $A_1$ and $A_2$. It should be emphasized that we are not aiming to train the diffractive network to implement the correct linear transformations for only a few input-output field pairs. Instead, despite the limited number of input/output field patterns used during the training process, our goal is to generalize to *any* pairs of ($i_1$, $o_1$) and ($i_2$, $o_2$) that satisfy $o_1 = A_1 i_1$ and $o_2 = A_2 i_2$. More details about the training data generation can be found in Methods.

To form two unique diffractive information processing pipelines in the same diffractive network for performing the linear transformations given by $A_1$ and $A_2$, as shown in Fig. 1a we matched the input fields and the diffractive output pairs, $\{(i_1, o'_1)\}$ and $\{(i_2, o'_2)\}$, with the input and output polarization channels of our diffractive system, i.e., $i_x = i_1$, $i_y = i_2$, $o_x = o'_1$ and $o_y = o'_2$. That is to say, the $A'_1$ transformation is performed by encoding the corresponding input field data $i_1$ into the x-polarized optical field within the input FOV, using e.g., an x-aligned linear polarizer, and decoding (sampling) the x-polarized component of the field within the output FOV as the computed output field $o'_1$ using e.g., an x-polarized analyzer. We denote this diffractive information processing channel as the channel ① in Fig. 1b. It is also a similar case for the $A'_2$ transformation, except this time the y-polarization is employed at the input and output FOVs, and this diffractive information processing channel is denoted as the channel ②. With this polarization encoding scheme, there are potentially two modes to perform the data inference through the same diffractive network: (1) in two sequential, successive accesses to the diffractive system, each time feeding the input data using its assigned polarization channel, and obtaining the corresponding output (see Fig. 1b, left); and (2) in single access to the diffractive system, by feeding the input data of both of the two polarization channels in parallel, and obtaining the two corresponding outputs simultaneously (see Fig. 1b, right). We term the former and latter approaches as the "sequential polarization access" (SeqPA) mode and the "simultaneous polarization access" (SimPA) mode, respectively, and analyze below these two modes of operation separately.

**2-channel polarization multiplexed all-optical diffractive computing using the sequential polarization access (SeqPA) mode**

As shown in Fig. 1b, left, with the input data $i_1$ and $i_2$ being separately and sequentially fed into the polarization channels ① and ②, respectively, the all-optical computed outputs $o'_1$ and $o'_2$ are also collected successively using the same diffractive network hardware. By employing this SeqPA strategy, we trained polarization multiplexed diffractive networks with different numbers of trainable diffractive neurons, i.e., $N = \{32^2, 44^2, 64^2, 92^2, 128^2, 180^2, 256^2\}$, all using the same training datasets $\{(i_1, o_1)\}$ and $\{(i_2, o_2)\}$ and the same number of epochs. To benchmark the performances of these multiplexed diffractive networks, for each transformation dataset and $N$, we also trained regular diffractive networks without the polarizer array or any polarization encoding/decoding at the input/output FOVs, which constitute our baseline. These regular diffractive networks, denoted as "*No pol.*" in our analyses, are trained to approximate *only one* linear transformation (i.e., either $A_1$ or $A_2$), and therefore they are referred to as $N_p = 1$ (no polarization multiplexing).

Figures 2b-e present the quantitative comparison of the all-optical transformation results obtained using the trained diffractive networks described above. Three different metrics were used to quantify the transformation accuracy and generalization performance of these diffractive networks: (1) the normalized transformation mean-squared error ($MSE_{\text{Transformation}}$), (2) the cosine similarity ($CosSim$) between the all-optical transforms and the target transforms, and (3) the mean-squared error between the diffractive network output fields and their ground truth ($MSE_{\text{Output}}$). These performance metrics are reported in Figs. 2b,c,d, as a function of the number of diffractive neurons ($N$) used in each design. Note that the transformation error of the polarization multiplexed diffractive systems is calculated per polarization channel. More details about the formulations of these performance metrics can be found in Methods. In Fig. 2b, it can be seen that the transformation errors of all the trained diffractive models monotonically decrease as $N$ increases, which is expected due to the increased degrees of freedom in the diffractive processor. In the standard diffractive networks without polarization multiplexing (dash-dotted curves labeled with "No pol. $A_1$" or "No pol. $A_2$"), the transformation errors for implementing $A_1$ or $A_2$ are almost the same (which indicates that these randomly selected matrices, $A_1$ and $A_2$, represent similar computational complexity; also see Fig. S1). The approximation errors of these standard diffractive networks, No pol. $A_1$ and No pol. $A_2$, both approach to 0 as $N$ approaches $N_i N_o = 64^2 \approx 4.1k$. In the polarization multiplexed diffractive models (solid curves labeled with "SeqPA ①" or "SeqPA ②"), the transformation errors $MSE_{\text{Transformation}}$ for the two distinct transforms computed through the two polarization channels are also very close to each other for all values of $N$, demonstrating no bias toward any specific polarization channel or transform. The approximation errors of these polarization multiplexed models approach to 0 as $N$ approaches $N_p N_i N_o = 2N_i N_o = 92^2 \approx 8.2k$. This finding indicates that compared with the baseline diffractive models that can only perform a *single* transform, performing two unique transforms using polarization multiplexing through the same diffractive model requires the number of trainable neurons $N$ to double. This conclusion is further supported by the results of the other two performance metrics, $CosSim$ (Fig. 2c) and $MSE_{\text{Output}}$ (Fig. 2d) that both show the same trends as in Fig. 2b: for the baseline diffractive models $CosSim$ and $MSE_{\text{Output}}$ approach 1 and 0 as $N$

approaches $N_iN_o$, respectively, while for the polarization multiplexed diffractive models, the two metrics approach 1 and 0 as $N$ approaches $N_pN_iN_o = 2N_iN_o$. Apart from the metrics that are used to evaluate the transformation performance, we also report the output diffraction efficiencies ($\eta$) of these diffractive models in Fig. 2e, which reveal that compared with the baseline diffractive networks (*No pol.*), the diffraction efficiencies of the polarization multiplexed diffractive models trained using the SeqPA mode reach a similar level.

To further demonstrate the performance of our polarization multiplexed diffractive networks, in Fig. 3 we show examples of the ground truth transformation matrices (i.e., $A_1$ and $A_2$) and their counterparts (i.e., $A'_1$ and $A'_2$) resulting from the diffractive designs with $N = \{44^2, 92^2, 180^2\}$, along with the amplitude and phase absolute errors. Exemplary complex-valued input-output fields from the same set of diffractive designs are also presented in Fig. 4. Figs. 3 and 4 reveal that for both of the polarization channels, when $N \geq N_pN_iN_o = 2N_iN_o$, the all-optical transformation matrices and the output complex fields very well match their ground truth targets with negligible absolute errors, which are also in line with the observations made in Fig. 2.

## 2-channel polarization multiplexed all-optical diffractive computing using the simultaneous polarization access (SimPA) mode

As an alternative to the sequential polarization access (SeqPA) used earlier, we also explored the use of the simultaneous polarization access (SimPA) mode in our all-optical computing framework. As shown in Fig. 1b, right, in single access to the diffractive system, the input complex-valued data $i_1$ and $i_2$ are fed into the polarization channels ① and ②, respectively, and the all-optical diffractive outputs $o'_1$ and $o'_2$ are collected at the same time through two orthogonal polarization states at the output FOV. Before we trained a new polarization multiplexed diffractive network from scratch using the SimPA mode, we first took our earlier diffractive designs trained using the SeqPA mode and tested them directly using the SimPA mode by inputting both polarization channels ① and ② *at the same time*, deviating from their training scheme, which only used SeqPA. The results of blindly testing the SeqPA-trained diffractive networks under the SimPA mode are shown in Fig. S2, which reveals inference results with significantly higher values of $MSE_{\text{Transformation}}$ and $MSE_{\text{Output}}$ and decreased values of $CosSim$, all of which indicate a performance degradation, when we operate a SeqPA-trained diffractive network using the SimPA mode. As shown in Fig. S3, this performance degradation is due to the "cross-talk" between the two transformation channels when both of the input polarization states are at the same time present, which was not considered during the SeqPA-based training process. These results highlight the necessity of training the diffractive system from scratch under the SimPA mode, so that the impact of this cross-talk can be taken into account and minimized during the iterative design process.

After training our diffractive models from scratch using the SimPA mode, we report their blind testing results in Fig. 5 using the solid curves labeled with "SimPA ①" and "SimPA ②". The results of the new diffractive designs trained using the SimPA mode demonstrate the success of all-optically performing two different linear transformations in parallel using polarization multiplexing. Our analysis (Fig. 5) also reveals the same conclusions discussed earlier for the models trained using the SeqPA mode: the all-optical transformation performance of polarization multiplexed diffractive networks very well match the ground truth, desired transformations as $N$ approaches $N_pN_iN_o = 2N_iN_o$.

We further compared the blind testing results of these two different modes of operation (SeqPA vs. SimPA) and performed a cross-talk field analysis (see Fig. S3). We found out that the amount of transformation cross-talk in the diffractive models trained using the SimPA mode (shown in the right column of Fig. S3c-d), is ~300-fold lower when compared with the amplitude values of the cross-talk observed in the diffractive designs trained using the SeqPA mode (shown in the left column of Fig. S3c-d). During the diffractive model training, these cross-talk fields are gradually eliminated (penalized) using the SimPA mode of operation to better approximate the ground truth fields. However, for the diffractive models trained under the SeqPA mode, such cross-talk fields are ignored (i.e., remain non-penalized during the training phase) since the SeqPA operation assumes successive access to the diffractive network, one input polarization state at a time. Stated differently, Seq-PA trained diffractive networks successfully approximate the target transformations only when they are tested under the same SeqPA mode of operation, and fail due to the field cross-talk when tested under the Sim-PA mode.

**4-channel polarization multiplexed all-optical diffractive computing**

So far, we have demonstrated to perform all-optical computing with 2-channel polarization multiplexing through a single diffractive network. To further exploit the polarization multiplexing capability of this diffractive computing framework, next, we explored a 4-channel polarization multiplexed design for performing 4 different arbitrarily-selected linear transformations through a single diffractive network (i.e., $N_p = 4$). Figure 6 illustrates the schematics of this framework. As depicted in Fig. 6b, by sequentially connecting one of the two input polarization states with one of the two output polarization states, four transformation channels, ①, ②, ③ and ④, can be formed to all-optically perform 4 distinct complex-valued transforms using the same diffractive processor. Compared to the 2-channel polarization multiplexed system reported earlier, the polarization states for the output field sampling in this 4-channel system are selected to be 45° and 135° linear polarization. This design choice is made to balance out the diffraction efficiencies of the resulting 4 different linear transformations that are all-optically performed by the diffractive network. Stated differently, this design choice introduces symmetry to all the input/output polarization combinations that are each assigned to a different linear transformation. In Fig. 6a and b, we denote the two output fields corresponding to the linear polarization directions at 45° and 135° as $\boldsymbol{o_\alpha}$ and $\boldsymbol{o_\beta}$, respectively.

In the light of our earlier findings that point to the need for more diffractive neurons in the case of $N_p = 2$ when compared to $N_p = 1$, here we employed 8 successive trainable diffractive layers to increase our degrees of freedom for $N_p = 4$ design (see Fig. 6a). Also, compared to the earlier 2-channel polarization multiplexed design, we included an additional linear polarizer array with the same configuration as before (with polarization orientations of 0°, 45°, 90° and 135°) to further enhance the spatial diversity of polarization modes within the diffractive processor. Same as the $N_p = 2$ diffractive designs, these linear polarizer arrays are pre-determined (i.e., non-trainable) and act as "polarization seeds" within the trained diffractive network.

Next, we generated random data to train and test our diffractive networks under $N_p = 4$. In addition to the two randomly-generated ground truth transforms $\boldsymbol{A_1}$ and $\boldsymbol{A_2}$ that were earlier used for the 2-channel models, we randomly generated two additional complex-valued transforms $\boldsymbol{A_3}$ and $\boldsymbol{A_4}$ and accordingly constructed the training and testing dataset consisting of the input and ground truth output fields. These four ground truth (target) transforms are visualized in Fig. 7a, and their differences can be found in Fig. S1. Following the training of the polarization multiplexed diffractive networks with different $N$, their transformation performance for $N_p = 4$ is analyzed in Fig. 7b-d based on the same set of performance metrics that were used earlier. These results reveal that, when $N$ approaches $N_p N_i N_o = 4 N_i N_o = 16.4k$, the $MSE_{\text{Transformation}}$ and $MSE_{\text{Output}}$ of all the four diffractive transformations approach 0, while the $CosSim$ approaches 1, demonstrating that all the target linear transformations ($A_1, A_2, A_3$ and $A_4$) can be successfully approximated by a single diffractive processor with a negligible error if $N \geq N_p N_i N_o$. This is the same conclusion that was reached earlier for $N_p = 2$.

To further demonstrate the success of these 4-channel polarization-multiplexed diffractive systems, in Fig. S4 we present the ground truth transformation matrices (i.e., $A_1, A_2, A_3$ and $A_4$) and their diffractive counterparts (i.e., $A'_1, A'_2, A'_3$ and $A'_4$) designed with $N$ = {14.3k, 66.5k}, along with the amplitude and phase errors made in each case. Furthermore, exemplary complex-valued output fields achieved by these diffractive systems are also shown in Fig. S5, all of which confirm the success of the presented 4-channel polarization-multiplexed diffractive designs. Finally, we also analyzed the output diffraction efficiencies of these diffractive models, reported in Fig. 7e. The results show that, compared to their counterparts without polarization encoding ($N_p = 1$), the polarization multiplexed diffractive models with $N_p = 4$ turn out to be less power efficient (per transformation), with an efficiency decrease of ~6 dB at the output FOV.

**Discussion**

Our results and analysis demonstrated that, using polarization multiplexing in a single diffractive network, one can all-optically perform a group of complex-valued arbitrary linear transformations at the same output FOV of the diffractive network. In practical applications, these different transformations can cover, for example, various machine vision tasks, such as detection, classification, and localization of objects, which can be programmed into different input/output polarization states. Compared to employing multiple diffractive subsystems, each one dedicated to performing a single task, integrating multiple tasks within the same diffractive system offers advantages of speed, versatility, compactness, and cost-effectiveness.

In addition to polarization multiplexing, we should note that other degrees of freedom can be used to implement multiple computational tasks through a single diffractive network. For example, one can divide the input/output FOVs of the diffractive network into multiple regions, where each region is assigned to a unique computing task through spatial-division multiplexing. It is also possible to achieve wavelength-division multiplexing by assigning different wavelengths or spectral bands to independent computing tasks and employing dispersive elements in the diffractive computing system. In contrast to these other possible methods of information multiplexing, the polarization-based multiplexing that we reported here requires solely the addition of linear polarizers to a diffractive network without changing its architecture. Such polarizers are readily available (e.g., polarizing films), even integrated with the individual pixels of polarization-based imaging systems[60], and can be adapted to a wide range of wavelengths. Furthermore, polarization multiplexing can be flexibly coupled with other multiplexing methods (such as spectral and/or spatial multiplexing) to further increase the computing capacity of the diffractive network.

Unlike the diffractive layers, where the transmission coefficients are trained and optimized to all-optically perform the target transformations, the design and arrangement of the seed polarizer arrays between the diffractive layers are treated as hyperparameters that are pre-determined and non-trainable. Optimization of the topology of such polarizer seeds within the diffractive volume could further enhance the approximation power of polarization multiplexed diffractive networks, which is left as future work.

We would like to also emphasize that the reported polarization-multiplexed diffractive networks can be directly applied to 2D arrays of phase and amplitude input data. Compared to other optical computing systems operating based on e.g., integrated photonics, which require 1D inputs and phase recovery if the information is represented in the phase channel, the capability to directly process and analyze raw 2D complex fields makes our framework highly advantageous for visual computing tasks. On the other hand, unless spatial light modulators (SLMs) are employed as part of the diffractive system (see e.g., the supplementary information of Ref. 23 for a discussion on reconfigurable networks), each physically fabricated diffractive network is fixed and would need to be retrained and fabricated again as the target transformations change, which is a limitation of *passive* diffractive systems.

There are additional limitations of the presented diffractive computing framework. First, polarization multiplexed diffractive computing systems present lower diffraction efficiencies at their output FOV compared to regular diffractive networks without polarization multiplexing. This is especially the case for the 4-channel polarization multiplexed system (see Fig. 7e). Several remedies can be used to improve the output diffraction efficiency such as e.g., adding a diffraction-efficiency-related penalty term to the training loss function, and/or restricting the diffractive layers to perform phase-only modulation. The efficacy of using these approaches in a regular diffractive network design (without polarization multiplexing) to improve the output diffraction efficiency has already been demonstrated in our earlier work[69]. To exemplify the performance of a phase-only diffractive design and how it can be used to improve the output diffraction efficiency, we trained phase-only diffractive networks from scratch for the 4-channel polarization multiplexing case ($N_p = 4$), the results of which are summarized in Fig. S6. This analysis revealed that phase-only diffractive designs can achieve significantly better output diffraction efficiencies (improved on average by ~12dB), while still successfully approximating the target linear transformations ($\boldsymbol{A_1}$, $\boldsymbol{A_2}$, $\boldsymbol{A_3}$ and $\boldsymbol{A_4}$). As a trade-off, however, these phase-only diffractive designs also exhibit reduced degrees of freedom compared to their complex-valued counterparts. As a result of this, we observed that all the target linear transformations were successfully approximated by a single phase-only diffractive processor when $N$ approached $2N_p N_i N_o = 8N_i N_o$. This 2-fold "threshold increase" in the number of diffractive features (i.e., $2N_p N_i N_o$ vs. $N_p N_i N_o$) is a direct reflection of the reduced number of trainable transmission parameters per diffractive layer due to the phase-only operation, which is a limitation of phase-only diffractive networks, despite their enhanced output diffraction efficiency. To further validate this conclusion, we also selected another set of 4 target linear transformations by changing the matrix elements to be real-valued, and used them as ground truth to train phase-only polarization multiplexed diffractive networks with $N_p = 4$. As shown in Fig. S7, our results reveal that these phase-only diffractive networks can successfully approximate the real-valued target linear transforms when $N \geq N_p N_i N_o = 4N_i N_o$, demonstrating a similar approximation performance, with significantly higher output diffraction efficiency compared to their complex-valued diffractive counterparts. These findings emphasize the value of phase-only diffractive network designs as a photon-efficient solution in polarization multiplexed diffractive computing, also providing an important rationale for planning the diffractive neuron budget ($N$) for a given computational task.

Another practical concern that needs to be discussed is the potential fabrication and alignment errors, surface reflections and material absorption within the diffractive network, which may altogether limit the performance and accuracy of diffractive computing. In our previous work,[36] we demonstrated that the performance degradation of a diffractive network caused by some of these experimental factors can be compensated by incorporating them as random variables into the physical forward model of the diffractive network during the training process. Some of these errors can also be mitigated by selecting appropriate fabrication methods, e.g., high-precision lithography, and using less absorptive materials. Moreover, our previous results[23,38,44,49,50] showed that these uncontrolled physical errors and imperfections did not lead to a significant discrepancy between the experimental and numerical, expected results, indicating the correctness of the assumptions involved in our optical forward model and training procedures.

In conclusion, we introduced a diffractive network-based all-optical computing framework that can perform multiple complex-valued, arbitrary linear transformations using polarization multiplexing. This framework is very compact; for instance, the system depicted in Fig. 1 has a total length of only 20λ in depth, where λ is the illumination wavelength. Our results show that when the number of diffraction elements/neurons, $N$, in a given diffractive network design approaches $N_p N_i N_o$, a group of $N_p$ arbitrarily-selected linear transforms can be all-optically computed at the output FOV of the network with negligible error. We believe that this polarization multiplexed diffractive computing framework can be used to build all-optical, passive processors that can execute multiple inference tasks in parallel. We further envision that artificially engineered materials with polarization manipulation capabilities (e.g., photonic crystals[70–72] and metamaterials[73,74]) can also be combined with advanced diffractive surface fabrication techniques (e.g., high-precision 3D additive manufacturing and photolithography) to allow the use of our diffractive computing framework in different parts of the electromagnetic spectrum.

**Materials and Methods**

**Forward model of the polarization multiplexed diffractive optical network.** Using Jones calculus[75], the complex-valued, polarization-multiplexed electrical field $\boldsymbol{E}$ at a spatial location $(x_m, y_m, z_m)$ can be represented as:

$$\boldsymbol{E}(x_m, y_m, z_m) = \begin{bmatrix} E_x(x_m, y_m, z_m) \\ E_y(x_m, y_m, z_m) \end{bmatrix} \quad (1).$$

In our implementation, $E_x$ and $E_y$ are computed in parallel throughout the entire diffractive system. Since the trainable diffractive layers are *not* polarization-sensitive, the complex-valued modulation generated by these thin diffractive layers is the same for the two orthogonal polarization states. The diffractive layers are assumed to be thin optical modulation elements, where the $m^{\text{th}}$ feature on the $k^{\text{th}}$ diffractive layer at location $(x_m, y_m, z_m)$ represents a complex-valued transmission coefficient, $t^k$, given by:

$$t^k(x_m, y_m, z_m) = a^k(x_m, y_m, z_m)\exp\left(j\phi^k(x_m, y_m, z_m)\right) \quad (2).$$

In Eq. 2, $a$ and $\phi$ denote the amplitude and phase coefficients, respectively. The amplitude and phase coefficients of the diffractive neurons, $a^k$ and $\phi^k$ ($k \in \{1, 2, \cdots, K\}$), are both trainable, with a permitted range of 0 to 1 and 0 to $2\pi$, respectively. Before the training starts, $a^k$ and $\phi^k$ are randomly initialized with a uniform ($U$) distribution of $U[0, 1]$ and $U[0, 2\pi)$, respectively. For a phase-only diffractive design $a^k = 1$. The size of each diffractive neuron on the transmissive layers and the width of the pixels of the input/output fields are both chosen as $\lambda/2$.

The diffractive layers are connected to each other by free-space wave propagation, which is modeled through the Rayleigh-Sommerfeld diffraction equation[23,32]:

$$w_m^k(x, y, z) = \frac{z - z_i}{r^2} \left( \frac{1}{2\pi r} + \frac{1}{j\lambda} \right) \exp\left( \frac{j2\pi r}{\lambda} \right) \tag{3},$$

where $w_m^k(x, y, z, \lambda)$ is the complex-valued field on the $m^{\text{th}}$ neuron of the $k^{\text{th}}$ layer at $(x, y, z)$ with a wavelength of $\lambda$, which can be viewed as a secondary wave generated from the source at $(x_m, y_m, z_m)$; and $r = \sqrt{(x - x_m)^2 + (y - y_m)^2 + (z - z_m)^2}$ and $j = \sqrt{-1}$. For the $k^{\text{th}}$ layer ($k \geq 1$, treating the input plane as the $0^{\text{th}}$ layer), the modulated optical field $E_p^k$ at location $(x_m, y_m, z_m)$ with a polarization state of $p$ ($p \in \{\text{x}, \text{y}\}$) is given by:

$$E_p^k(x_m, y_m, z_m) = t^k(x_m, y_m, z_m) \cdot \sum_{n \in S} E_p^{k-1}(x_n, y_n, z_n) \cdot w_m^{k-1}(x_m, y_m, z_m) \tag{4},$$

where $S$ denotes all the pixels on the previous diffractive layer. For all the diffractive networks trained in this paper, the axial distances $d_0, d_1, \ldots, d_K$ are all chosen as $4\lambda$.

When modeling the polarizer elements in our diffractive system, we used Jones matrices to represent the modulation of the complex field brought by the input polarizer, output analyzer, or the polarizer array at location $(x, y, z)$, the process of which can be written as:

$$\boldsymbol{E}_{\text{out}}(x, y, z) = \boldsymbol{J}_{\text{linear}}(x, y, z) \boldsymbol{E}_{\text{in}}(x, y, z) \tag{5},$$

where $E_{\text{in}}$ and $E_{\text{out}}$ are the vectors denoting the input and output complex field before and after the polarization modulation, each containing two orthogonal components along the x and y directions, i.e., $\boldsymbol{E}_{\text{out}}(x, y, z) = \begin{bmatrix} E_{\text{out},x}(x, y, z) \\ E_{\text{out},y}(x, y, z) \end{bmatrix}$ and $\boldsymbol{E}_{\text{in}}(x, y, z) = \begin{bmatrix} E_{\text{in},x}(x, y, z) \\ E_{\text{in},y}(x, y, z) \end{bmatrix}$.

$\boldsymbol{J}_{\text{linear}}(x, y, z)$ represents the Jones matrix of a linear polarizer element, which is given by:

$$\boldsymbol{J}_{\text{linear}}(x, y, z) = \begin{bmatrix} \cos^2\theta(x, y, z) & \cos\theta(x, y, z) \sin\theta(x, y, z) \\ \sin\theta(x, y, z) \cos\theta(x, y, z) & \sin^2\theta(x, y, z) \end{bmatrix} \tag{6},$$

where $\theta(x, y, z)$ is the angle between the x-axis and the polarizing axis of the linear polarizer located at $(x, y, z)$. For the non-trainable, pre-determined polarizer array that is composed of multiple square-shaped linear polarizers, we used in total 4 types of linear polarizer units with 4 different polarizing axis directions, $\theta = \{0, 0.25\pi, 0.5\pi, \text{and } 0.75\pi\}$. As illustrated in Fig. 1a, these 4 different types of linear polarizers are spatially binned to have a 2×2 period and repeated with 3 periods in each direction, extending into a square region. The side length of each linear polarizer array is 24λ. The residual space surrounding the polarizer array is filled with air, without any polarization modulation. For all the diffractive network designs presented in this paper, the axial distances (i.e., $d_p$, $d_{p1}$ and $d_{p2}$) between the pre-determined polarizer arrays and the adjacent diffractive layers in front of them are all empirically chosen as 0; stated differently, each linear polarizer array is attached to the isotropic diffractive layer in front of it.

**Preparation of the linear transformation datasets.** In our diffractive network designs, the input and output FOVs have the same size of 8 × 8 pixels, i.e., $i_c, o_c \in \mathbb{C}^{8 \times 8}$ ($c \in \{1, 2, 3, 4\}$). The size of the transformation matrices is equal to 64 × 64, i.e., $A_c \in \mathbb{C}^{64 \times 64}$ ($c \in \{1, 2, 3, 4\}$). The amplitude and phase components of the complex-valued transformation matrices $A_c$ used in this paper were generated with a uniform ($U$) distribution of $U[0, 1]$ and $U[0, 2\pi)$, respectively, using the pseudo-random number generation function *random.uniform()* built-in NumPy. Different random seeds were used to generate these transformation matrices to ensure they were uniquely different (see Fig. S1). Next, the amplitude and phase components of the input fields $i_c$ ($c \in \{1, 2, 3, 4\}$) were also randomly generated with a uniform ($U$) distribution of $U[0, 1]$ and $U[0, 2\pi)$, respectively. The ground truth (target) fields $o_c$ ($c \in \{1, 2, 3, 4\}$) were generated by calculating $o_c = A_c i_c$. For each $A_c$ ($c \in \{1, 2, 3, 4\}$) we generated a total of 70,000 input/output complex fields to form a dataset, divided into three parts: training, validation, and testing, each containing 55,000, 5,000, and 10,000 complex-valued field pairs, respectively.

**Training loss function.** For training of our diffractive networks, we used the mean-squared-error (MSE) loss function, which is defined as:

$$\mathcal{L}_{\text{MSE},c} = E\left[\frac{1}{N_o}\sum_{n=1}^{N_o}|\widehat{o_1}[n] - \widehat{o_1'}[n]|^2\right]$$

$$= E\left[\frac{1}{N_o}\sum_{n=1}^{N_o}|\sigma_c o_c[n] - \sigma_c' o_c'[n]|^2\right] \quad (7),$$

where $E[\cdot]$ denotes the average across the current batch, $c$ stands for the $c^{\text{th}}$ polarization channel that is being accessed, and $[n]$ indexes the $n^{\text{th}}$ element of the vector. $\sigma_c$ and $\sigma_c'$ are the coefficients used to normalize the energy of the ground truth (target) field $o_c$ and the diffractive-network output field $o_c'$, respectively, which are given by:

$$\sigma_c = \frac{1}{\sqrt{\sum_{n=1}^{N_o}|o_c[n]|^2}} \quad (8),$$

$$\sigma_c' = \frac{\sum_{n=1}^{N_o}\sigma_c o_c[n] o_c'^*[n]}{\sum_{n=1}^{N_o}|o_c'[n]|^2} \quad (9).$$

During the training of the diffractive networks using the SeqPA mode, each polarization channel of the diffractive network is accessed and evaluated cyclically based on the order of the channel number. For instance, for the 2-channel polarization multiplexed design illustrated in Fig. 1b, left, the access sequence during the training is set to be {①, ②, ①, ②, ...}; for the 4-channel polarization multiplexed design illustrated in Fig. 6, the access sequence is {①, ②, ③, ④, ①, ②, ③, ④, ...}. During the access of a certain polarization channel, the diffractive network is fed with one batch of the training input/output complex fields corresponding to the transformation matrix assigned to this channel, and then trained based on the average loss across this batch. Thus, the loss function for training the diffractive designs through the $c^{th}$ polarization channel using the SeqPA mode, $\mathcal{L}_{\text{Seq},c}$, can be simply written as:

$$\mathcal{L}_{\text{Seq},c} = \mathcal{L}_{\text{MSE},c} \qquad (10).$$

During the training of the diffractive networks using the SimPA mode, as illustrated in Fig. 1b, right, all the polarization channels of the diffractive network are accessed simultaneously, and the training data are fed into the channels at the same time. For this SimPA mode, the diffractive network is trained based on the loss averaged across the different polarization channels and complex-valued fields in the current batch, where the loss function $\mathcal{L}_{\text{Sim}}$ can be written as:

$$\mathcal{L}_{\text{Sim}} = \frac{1}{N_p}\sum_{c=1}^{N_p} \mathcal{L}_{\text{MSE},c} \qquad (11).$$

**Performance metrics used for the quantification of all-optical transformation errors.** To quantitatively evaluate the transformation results of the polarization-multiplexed diffractive networks, four performance metrics were calculated per polarization channel of the diffractive designs using the testing data set: (1) the normalized transformation mean-squared error ($MSE_{\text{Transformation}}$), (2) the cosine similarity ($CosSim$) between the all-optical transforms and the target transforms, (3) the normalized mean-squared error between the diffractive network output fields and their ground truth ($MSE_{\text{Output}}$), and (4) the output diffraction efficiency ($\eta$). The transformation error for the $c^{th}$ polarization channel of the diffractive network, $MSE_{\text{Transformation},c}$, is defined as:

$$\begin{aligned} MSE_{\text{Transformation},c} &= \frac{1}{N_i N_o} \sum_{n=1}^{N_i N_o} |\boldsymbol{a}_c[n] - m_c \boldsymbol{a}'_c[n]|^2 \\ &= \frac{1}{N_i N_o} \sum_{n=1}^{N_i N_o} |\boldsymbol{a}_c[n] - \widehat{\boldsymbol{a}'_c}[n]|^2 \end{aligned} \qquad (12),$$

where $\boldsymbol{a}_c$ is the vectorized version of the ground truth transformation matrix assigned to the $c^{th}$ polarization channel $\boldsymbol{A}_c$, i.e., $\boldsymbol{a}_c = \text{vec}(\boldsymbol{A}_c)$. $\boldsymbol{a}'_c$ are the vectorized version of $\boldsymbol{A}'_c$, which is the all-optical transformation matrix computed using the optimized diffractive transmission coefficients. $m_c$ is a scalar normalization coefficient used to eliminate the effect of diffraction-efficiency related scaling mismatch between $\boldsymbol{A}_c$ and $\boldsymbol{A}'_c$, i.e.,

$$m_c = \frac{\sum_{n=1}^{N_i N_o} \boldsymbol{a}_c[n]\boldsymbol{a}'^*_c[n]}{\sum_{n=1}^{N_i N_o} |\boldsymbol{a}'_c[n]|^2} \qquad (13).$$

The cosine similarity between the all-optical transform and their target transform for the $c^{th}$ polarization channel, $CosSim_c$, is defined as:

$$CosSim_c = \frac{|\boldsymbol{a}_c^H \widehat{\boldsymbol{a}}_c'|}{\sqrt{\sum_{n=1}^{N_i N_o}|\boldsymbol{a}_c[n]|^2} \sqrt{\sum_{n=1}^{N_i N_o}|\widehat{\boldsymbol{a}_c'}[n]|^2}} \quad (14).$$

The normalized mean-squared error between the diffractive network outputs and their ground truth for the $c^{th}$ polarization channel, $MSE_{Output,c}$, is defined using the same formula as in Eq. 7 (the loss function used during the training process), except for that $E[\cdot]$ is calculated across the entire testing set.

The mean diffraction efficiency $\eta_c$ for the $c^{th}$ polarization channel of the diffractive system is defined as:

$$\eta_c = E\left[\frac{\sum_{n=1}^{N_o}|\boldsymbol{o}_c'[n]|^2}{\sum_{n=1}^{N_i}|\boldsymbol{i}_c[n]|^2}\right] \quad (15).$$

**Training-related details.** All the diffractive optical networks used in this work were simulated and trained using Python (v3.8.11) and TensorFlow (v2.6.0, Google Inc.). We selected Adam optimizer[76] for training all the models, and its parameters were taken as the default values in TensorFlow and kept identical in each model. The batch size and learning rate were set as 8 and 0.001, respectively. The training of the diffractive network models using the SimPA mode was performed with 50 epochs. For training the diffractive models using the SeqPA mode, the 2-channel and 4-channel polarization multiplexed designs were trained for 100 and 200 epochs, respectively, so that equivalently 50 epochs are dedicated for training each polarization channel of these designs. The best models were selected based on the MSE loss calculated on the validation data set. For the training of our diffractive models, we used a desktop computer with a GeForce GTX 1080Ti graphical processing unit (GPU, Nvidia Inc.) and Intel® Core™ i7-8700 central processing unit (CPU, Intel Inc.) and 64 GB of RAM, running Windows 10 operating system (Microsoft Inc.). The typical time to train a diffractive network model using the SeqPA mode with 2 and 4 polarization channels is ~7 and ~14 hours, respectively. The training time for a diffractive model using the SimPA mode with 2 polarization channels is ~4 hours.

**Supplementary Materials:** This file contains Supplementary Figures S1-S7.

# Figures

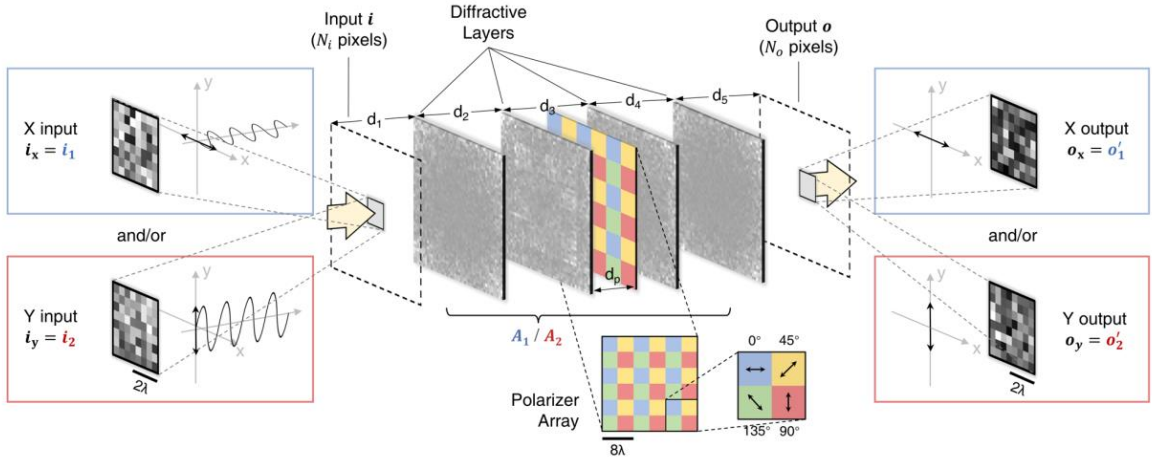

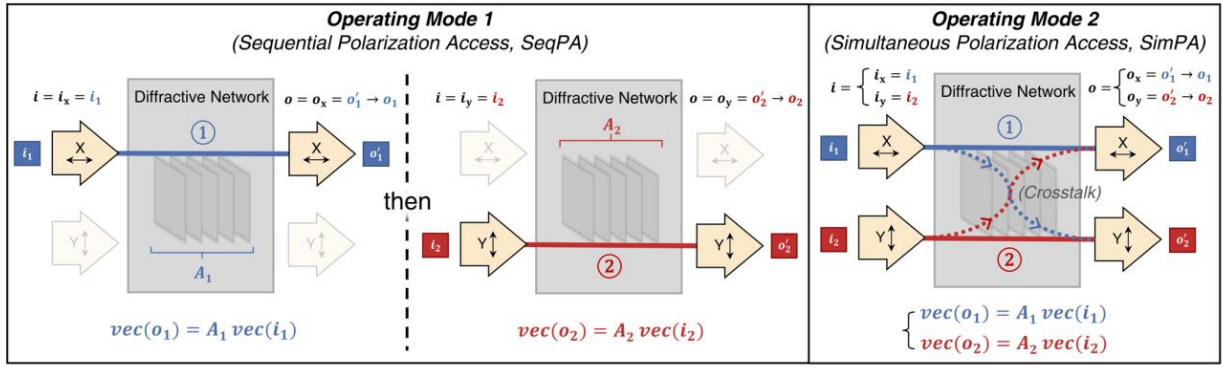

**Figure 1. Schematic of polarization multiplexed all-optical diffractive computing. a**, Optical layout of the polarization-encoded diffractive network, where four isotropic diffractive layers and one array of linear polarizers are jointly used to perform two distinct, complex-valued linear transformations between the input field $i$ and the output field $o$ by using polarization encoding/decoding at the input/output FOVs. **b**, Schematic for the sequential polarization access (SeqPA, left) mode and the simultaneous polarization access (SimPA, right) mode that can be used to operate the 2-channel polarization multiplexed diffractive network.

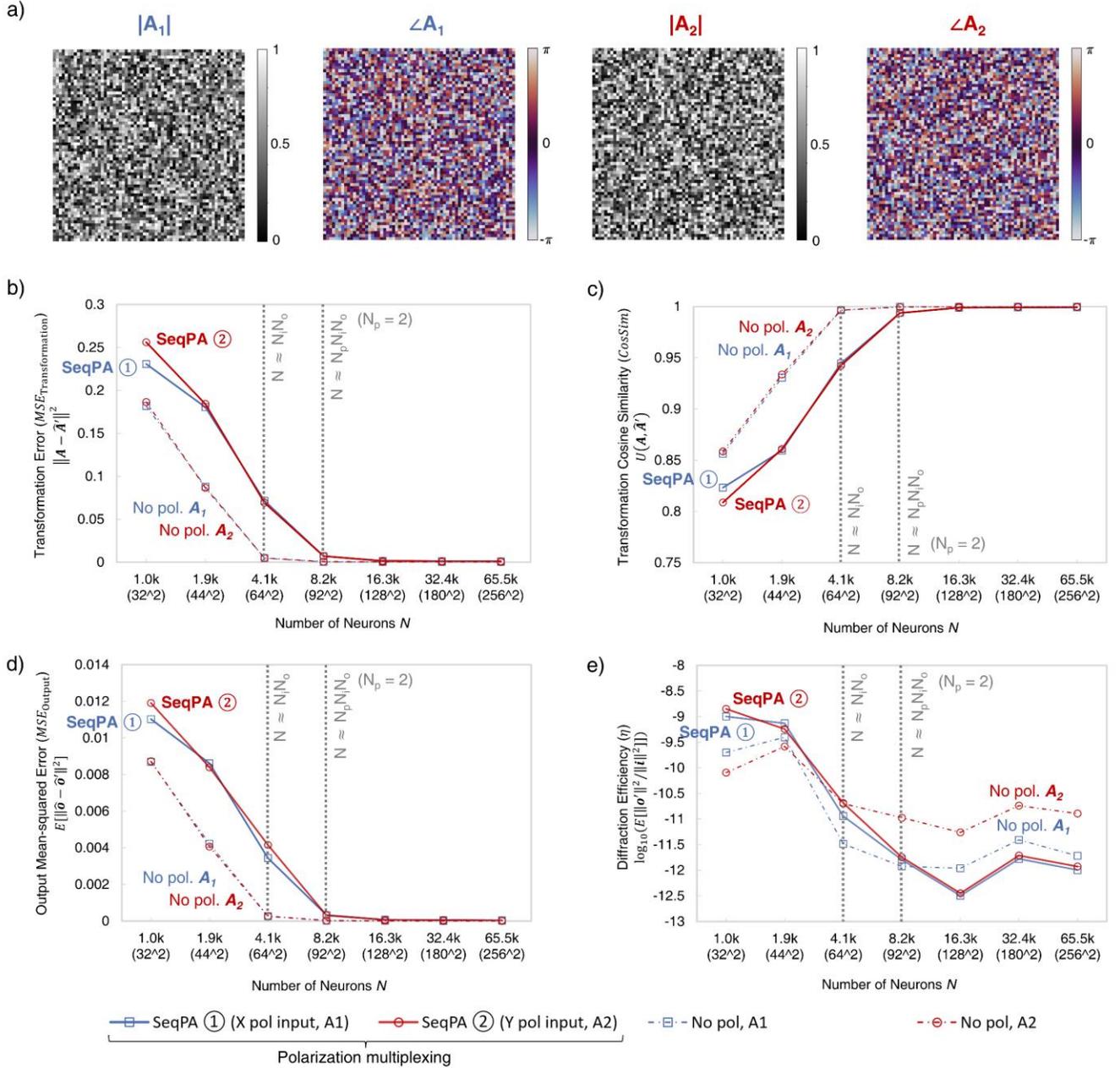

**Figure 2. Diffractive all-optical transformation results for 2-channel polarization multiplexing using the sequential polarization access (SeqPA) mode. a**, Amplitude and phase of the arbitrarily generated matrices $A_1$ and $A_2$, which serve as the ground truth (target) for the diffractive all-optical transformations. **b**, Curves representing the normalized mean-squared error between the ground truth transformation matrices ($A_1$ and $A_2$) and the all-optical transforms ($A'_1$ and $A'_2$) resulting from the trained diffractive networks as a function of the number of diffractive neurons $N$. The solid curves are achieved by the polarization multiplexed diffractive networks trained using the SeqPA mode, which are compared with the dashed curves achieved by the regular diffractive networks trained with the same set of $N$ but without any polarization multiplexing. For the polarization multiplexed models, the results for the two polarization channels ① and ②, corresponding to transforms $A'_1$ and $A'_2$, are shown in separate curves that are labeled with "SeqPA

①" and "SeqPA ②", respectively. For the regular diffractive models without polarization multiplexing, the results for all-optical approximation of $A_1$ and $A_2$ are shown in separate curves labeled with "No pol. $A_1$" and "No pol. $A_2$", respectively. The space between the simulation data points is linearly interpolated. **c**, Same as (b) but the cosine similarity between the all-optical transforms and their ground truth shown in (a) is reported. **d**, Same as (b) but the mean-squared error between the diffractive network output fields and their ground truth is reported. **e**, Diffraction efficiency of the presented diffractive networks.

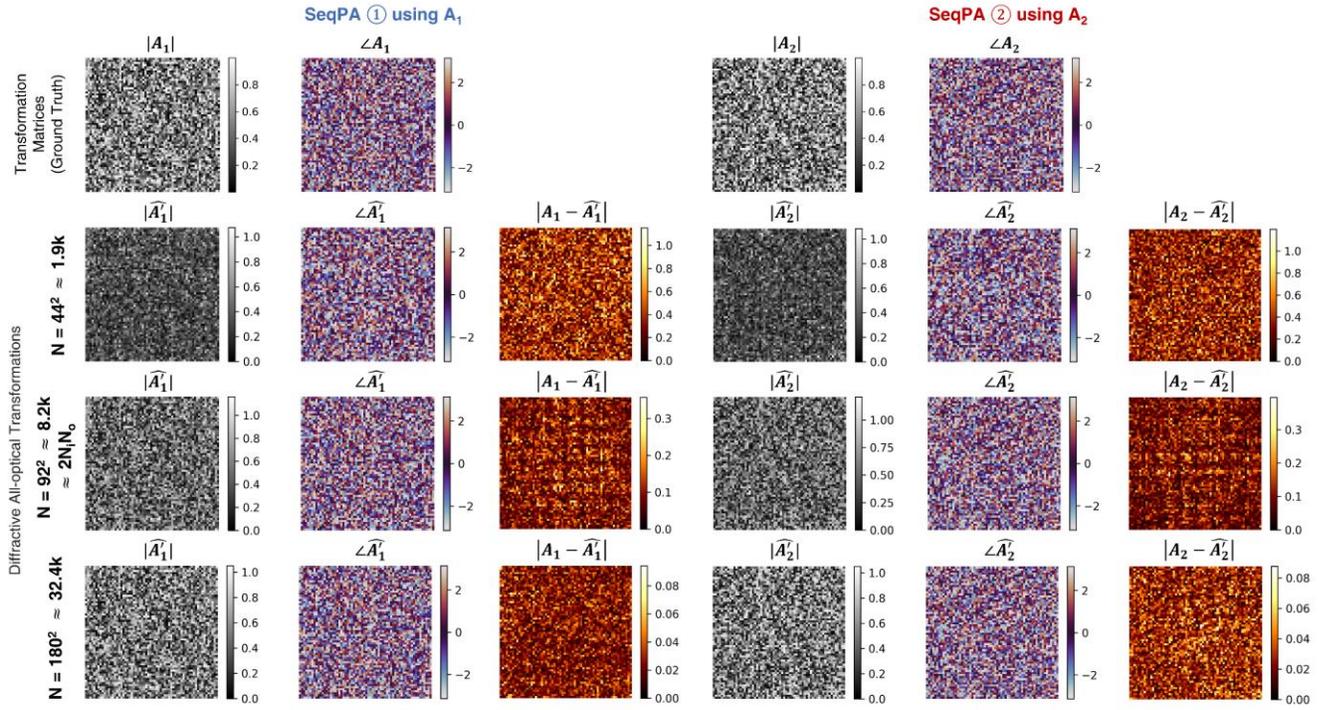

**Figure 3.** All-optical transformation matrices estimated by the 2-channel polarization multiplexed diffractive designs trained using the SeqPA mode with $N = 44^2$, $92^2$ and $180^2$, and their differences from the ground truth matrices.

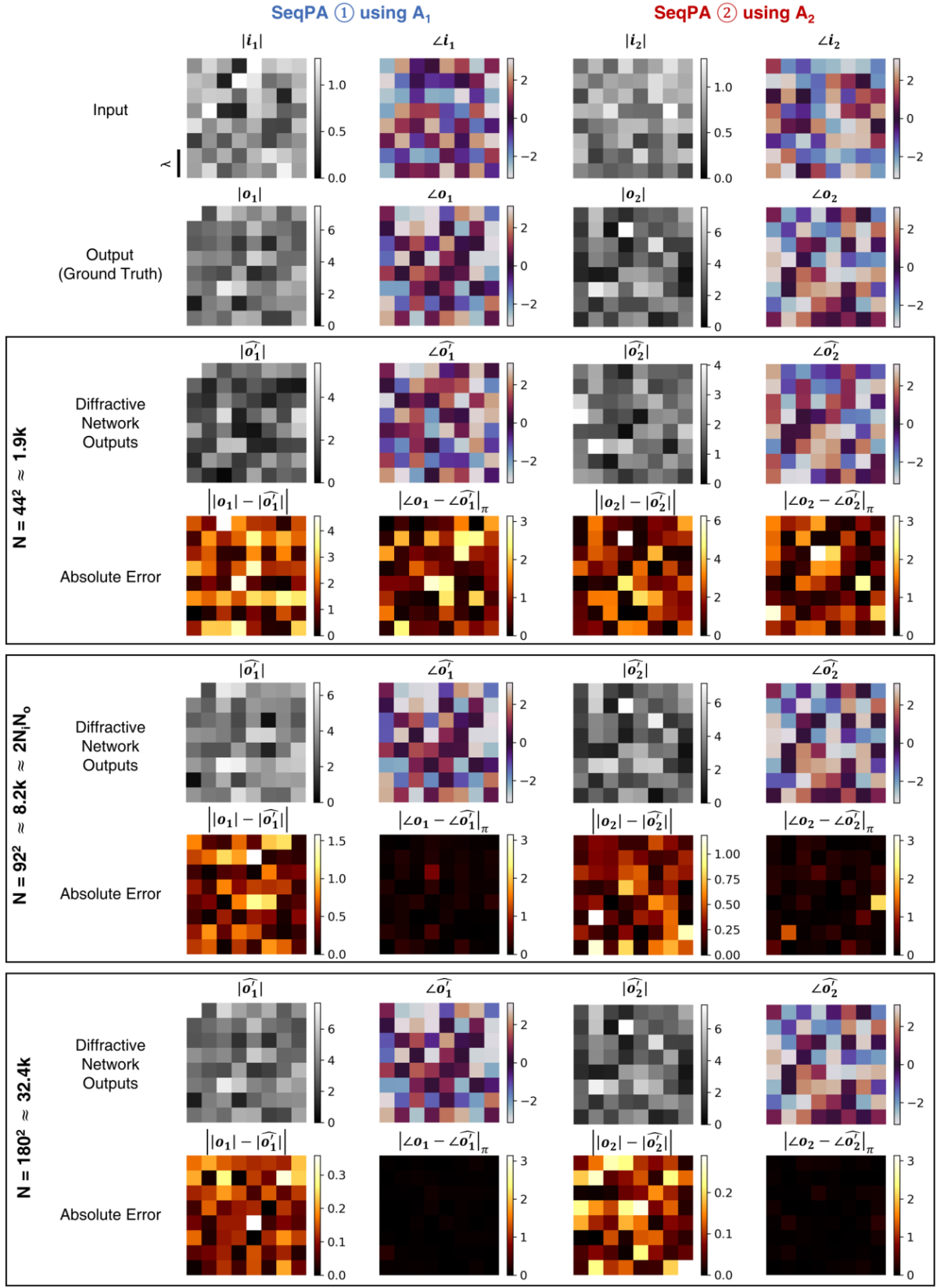

**Figure 4. Examples of input/output complex fields for the ground truth transformations**

presented in Figures 2 and 3 along with the output fields computed by the 2-channel polarization multiplexed diffractive designs trained with the SeqPA mode using $N = 44^2$, $92^2$ and $180^2$. Note that $\left|\angle o - \angle \widehat{o'}\right|_\pi$ indicates the wrapped phase difference between the ground truth output field $o$ and the normalized diffractive network output field $\widehat{o'}$

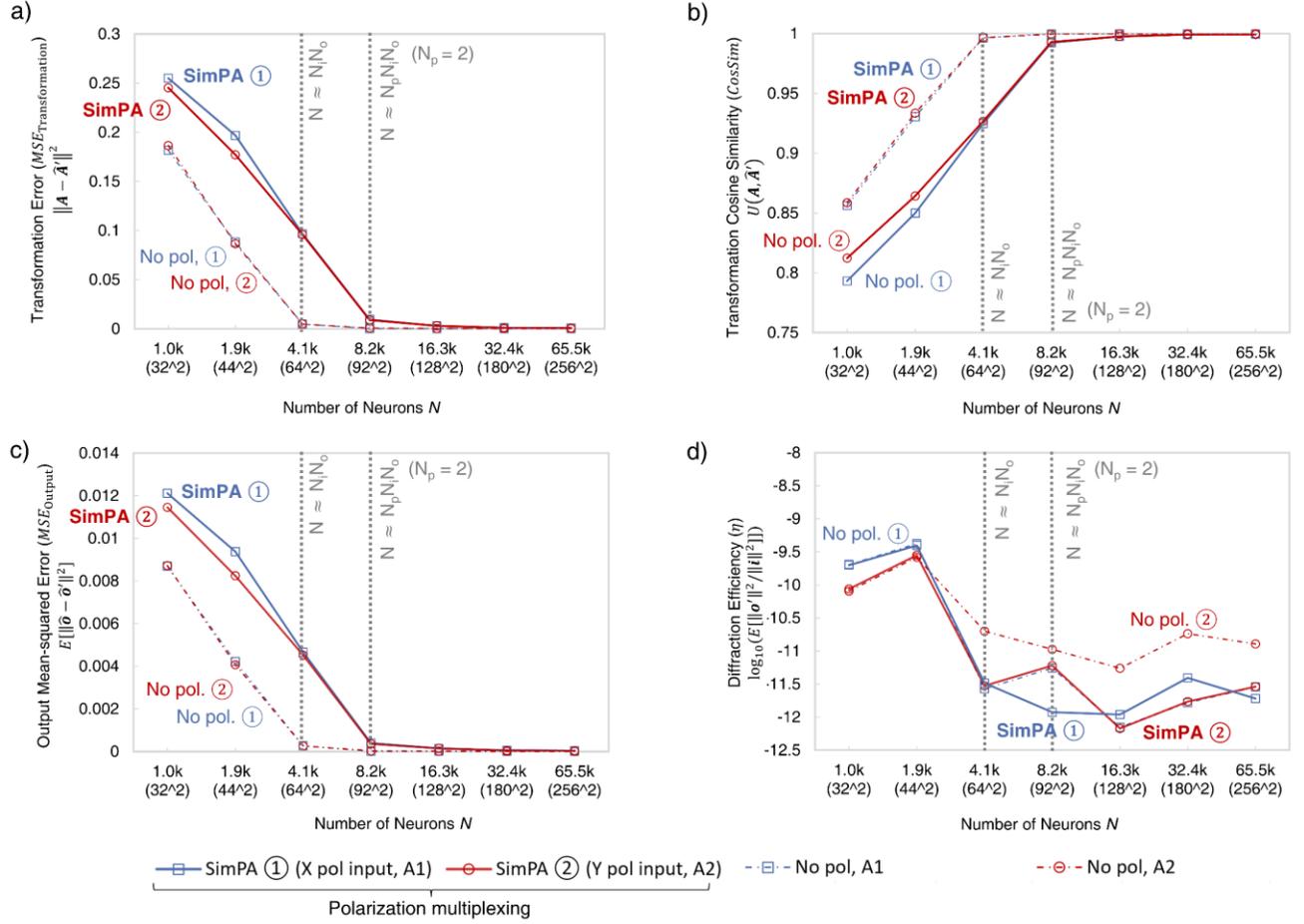

**Figure 5. Diffractive all-optical transformation results for 2-channel polarization multiplexing using the simultaneous polarization access (SimPA) mode. a**, Curves representing the normalized mean-squared error between the ground truth transformation matrices ($A_1$ and $A_2$) and the all-optical transforms ($A'_1$ and $A'_2$) resulting from the trained diffractive networks as a function of the number of diffractive neurons $N$. The solid curves are achieved by the polarization multiplexed diffractive systems trained using the SimPA mode, which are compared with the dashed curves achieved by the regular diffractive networks trained with the same set of $N$ but without any polarization multiplexing. For the polarization multiplexed models, the results for the two polarization channels ① and ②, corresponding to transforms $A'_1$ and $A'_2$, are shown in separate curves that are labeled with "SimPA ①" and "SimPA ②", respectively. For the regular models without polarization multiplexing, the results for all-optical approximation of $A_1$ and $A_2$ are shown in separate curves labeled with "No pol. $A_1$" and "No pol. $A_2$", respectively. The space between the simulation data points is linearly interpolated. **b**, Same as (a) but the cosine similarity between the all-optical transforms and their ground truth is reported. **c**, Same as (a) but the mean-squared error between the diffractive network output fields and their ground truth is reported. **d**, Diffraction efficiency of the presented diffractive networks.

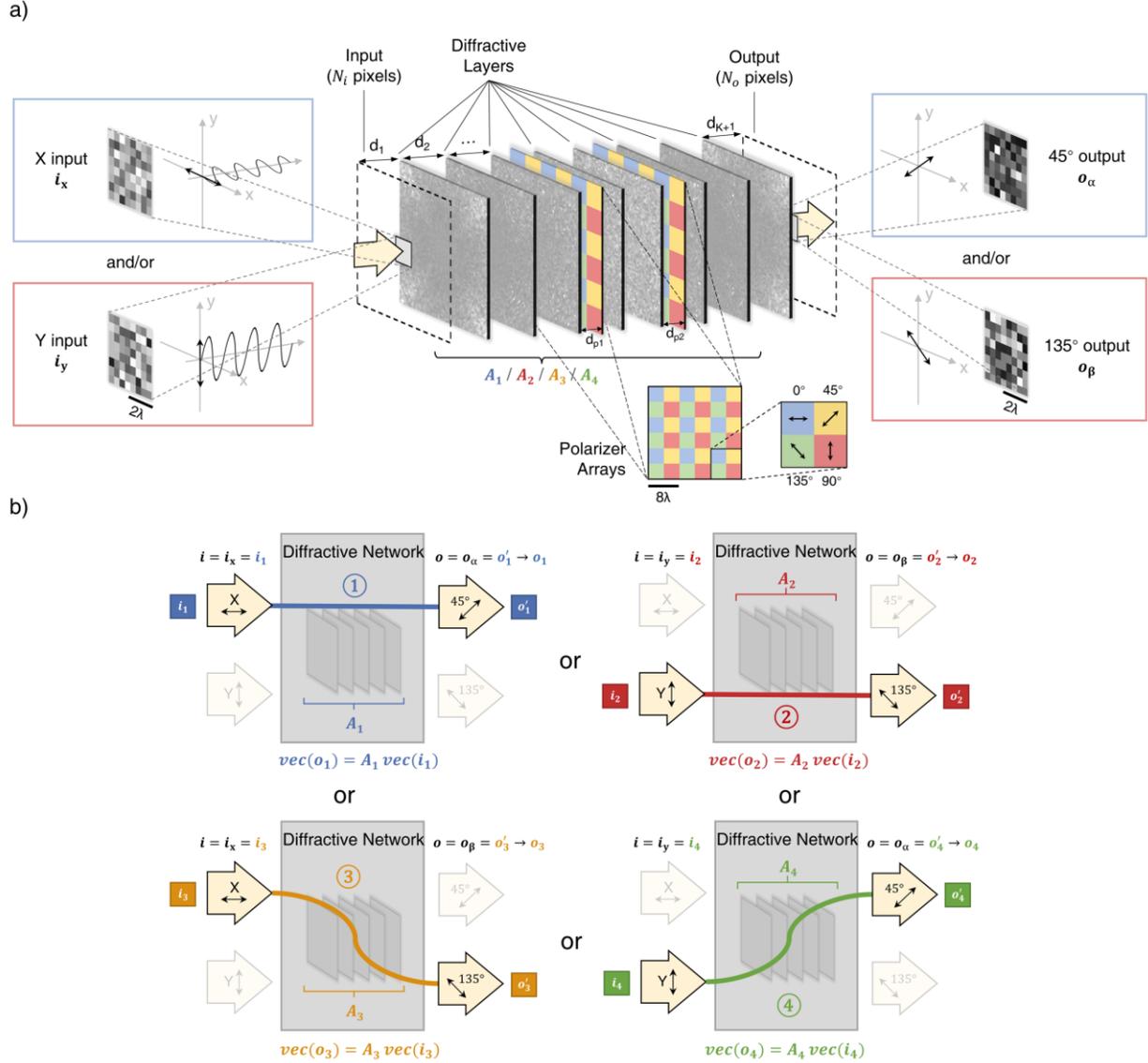

**Figure 6. Schematic of 4-channel polarization multiplexed all-optical diffractive computing framework for performing four unique linear transformations through a single diffractive network. a**, Optical layout of the polarization-encoded diffractive network, where eight trained diffractive layers and two arrays of linear polarizers are jointly used to perform four distinct, complex-valued linear transformations between the input field ***i*** and the output field ***o*** by using polarization encoding/decoding at the input/output FOVs. **b**, Schematic for the operation of the 4-channel polarization multiplexed all-optical computing framework, where the four polarization channels, ①, ②, ③ and ④, are formed by sequentially connecting one of the two input polarization states with one of the two output polarization states.

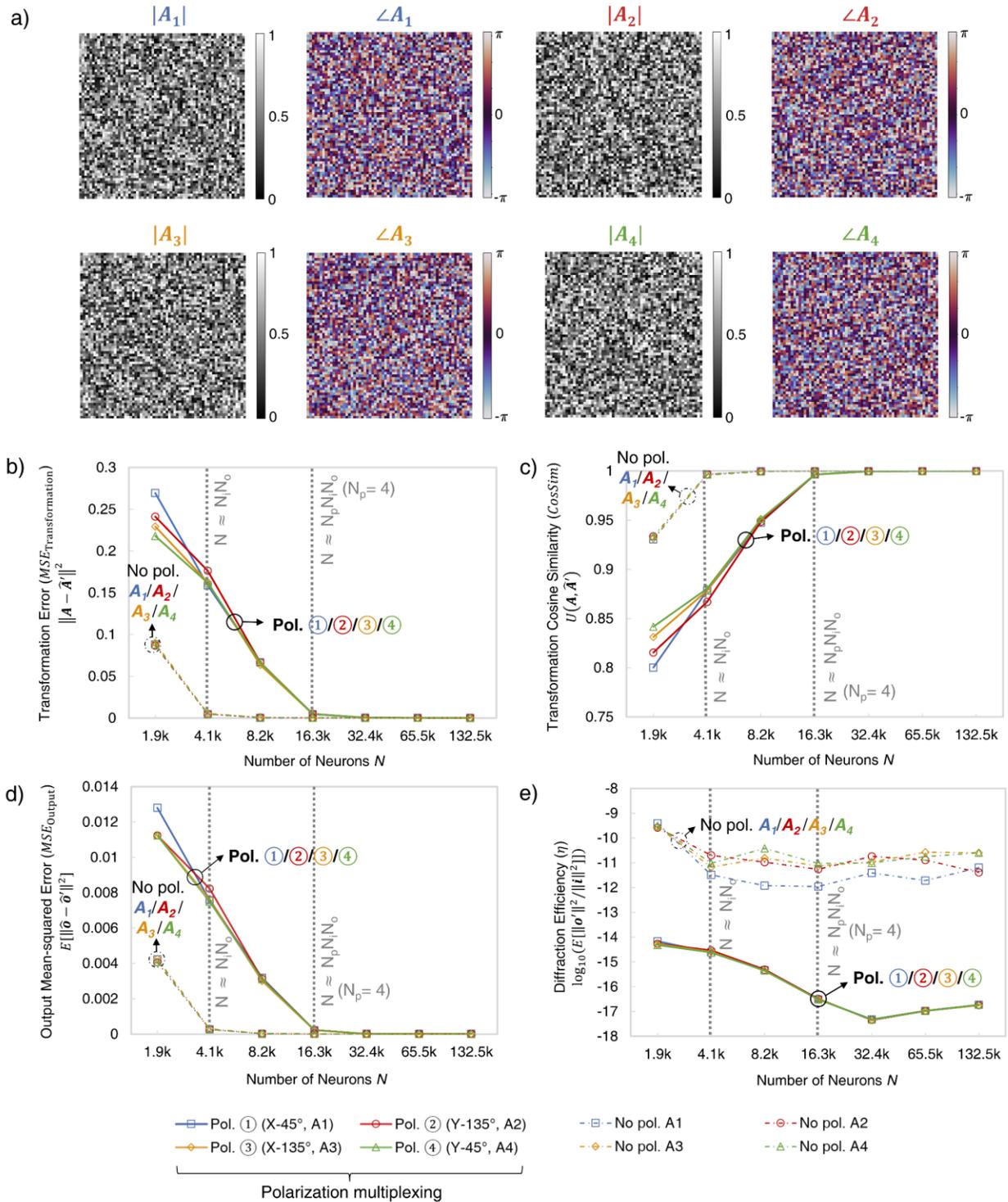

**Figure 7. Diffractive all-optical transformation results for 4-channel polarization multiplexing of four distinct arbitrary linear transforms (depicted in Fig. 6). a**, Amplitude and phase of the arbitrarily generated matrices $A_1$, $A_2$, $A_3$ and $A_4$, which serve as the ground truth (target) for the diffractive all-optical transformations. **b**, Curves representing the normalized mean-squared error between the ground truth transformation matrices ($A_1$, $A_2$, $A_3$ and $A_4$) and the all-

optical transforms ($A'_1$, $A'_2$, $A'_3$ and $A'_4$, examples shown in Fig. S4) resulting from the trained diffractive networks as a function of the number of diffractive neurons $N$. The solid curves are achieved by the 4-channel polarization multiplexed diffractive systems, which are compared with the dashed curves achieved by the regular diffractive networks trained with the same set of $N$ but without polarization multiplexing. For the polarization multiplexed models, the results for the four polarization channels ①, ②, ③ and ④ are shown in separate curves but jointly labeled with "Pol. ①/②/③/④" due to the spatial overlap of these curves. For the regular diffractive models without polarization multiplexing, the results for all-optical approximation of $A_1$, $A_2$, $A_3$ and $A_4$ (individually) are shown in separate curves but jointly labeled with "No pol. $A_1/A_2/A_3/A_4$" due to the spatial overlap of these curves. The space between the simulation data points is linearly interpolated. **c**, Same as (b) but cosine similarity between the all-optical transforms and their ground truth is reported. **d**, Same as (b) but the mean-squared error between the diffractive network output fields and their ground truth is reported. **e**, Diffraction efficiency of the presented diffractive networks.